\documentclass[[final,3p,times,twocolumn]{elsarticle}
\usepackage{graphicx}
\usepackage{dcolumn}
\usepackage{bm}
\usepackage{amsmath}
\usepackage{comment}
\usepackage{epstopdf} 
\usepackage{xr-hyper}
\usepackage{lineno,hyperref}
\usepackage{color} 
\usepackage{soul} 

\makeatletter
\newcommand*{\addFileDependency}[1]{
	\typeout{(#1)}
	\@addtofilelist{#1}
	\IfFileExists{#1}{}{\typeout{No file #1.}}
}
\makeatother

\usepackage{amssymb}

\usepackage{lineno,hyperref}
\modulolinenumbers[5]

\journal{Ultramicroscopy}

\begin{document}

\begin{frontmatter}



\title{Event driven 4D STEM acquisition with a Timepix3 detector: microsecond dwell time and faster scans for high precision and low dose applications}

\author[emat,nanolab]{D. Jannis}
\author[emat,nanolab]{C. Hofer}
\author[emat,nanolab]{C. Gao}
\author[emat,nanolab]{X. Xie}
\author[emat,nanolab]{A. B\'ech\'e}
\author[emat,nanolab]{T.J. Pennycook}
\author[emat,nanolab]{J. Verbeeck\corref{mycorrespondingauthor}}
\cortext[mycorrespondingauthor]{Corresponding author}
\ead{jo.verbeeck@uantwerp.be}

\address[emat]{EMAT, University of Antwerp, Groenenborgerlaan 171, 2020 Antwerp, Belgium}
\address[nanolab]{NANOlab Center of Excellence, University of Antwerp, Groenenborgerlaan 171, 2020 Antwerp, Belgium}

\begin{abstract}
	Four dimensional scanning transmission electron microscopy (4D STEM) records the scattering of electrons in a material in great detail. The benefits offered by 4D STEM are substantial, with the wealth of data it provides facilitating for instance high precision, high electron dose efficiency phase imaging via center of mass or ptychography based analysis. However the requirement for a 2D image of the scattering to be recorded at each probe position has long placed a severe bottleneck on the speed at which 4D STEM can be performed. Recent advances in camera technology have greatly reduced this bottleneck, with the detection efficiency of direct electron detectors being especially well suited to the technique. However even the fastest frame driven pixelated detectors still significantly limit the scan speed which can be used in 4D STEM, making the resulting data susceptible to drift and hampering its use for low dose beam sensitive applications. Here we report the development of the use of an event driven Timepix3 direct electron camera that allows us to overcome this bottleneck and achieve 4D STEM dwell times down to 100~ns; orders of magnitude faster than what has been possible with frame based readout. We characterise the detector for different acceleration voltages and show that the method is especially well suited for low dose imaging and promises rich datasets without compromising dwell time when compared to conventional STEM imaging.

\end{abstract}



\begin{keyword}
	Scanning transmission electron microscopy \sep 4D STEM \sep Low dose \sep Event based detection
	
	
	
\end{keyword}

\end{frontmatter}



\section{Introduction}
\begin{figure*}[!htb]
\includegraphics[width=\textwidth]{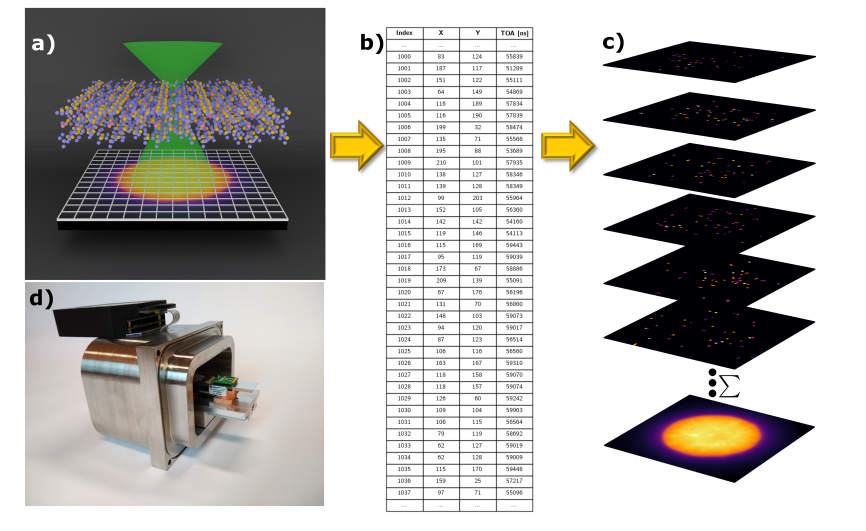}
\caption{\label{fig:setup} \textbf{(a)} The schematic setup, where a convergent electron beam is scanned across the sample and during the scan, the point of impact and TOA is measured for each incoming electron. \textbf{(b)} The data list created by the detector where for every event the point of impact and TOA is indicated. \textbf{(c)} When the probe position at each time is accurately known, a diffraction pattern belonging to the time range for a given probe position can be obtained resulting in a full 4D STEM dataset. \textbf{(d)} Custom built retractable assembly to place the Timepix3 detector in the electron microscope column.  
}	
\end{figure*}

The development of direct pixelated detectors has revolutionized the imaging capabilities of multiple scientific fields, from particle tracking to X-ray and electron microscopy \cite{procz_x-ray_2019, bergmann_3d_2019}. In electron microscopy (EM), the efficiency of direct detectors has provided a leap forward in the dose efficiency possible in transmission electron microscopy (TEM). For fields such as molecular biology the resolution provided by EM is limited more by the poor signal achievable using the extremely low doses required to avoid unacceptable damage to the material than by the electron optics themselves. For such fields therefore dose efficiency is of the upmost importance, and the greatly enhanced dose efficiency provided by direct electron detectors made possible the so called resolution revolution in cryo-EM \cite{kuhlbrandt_resolution_2014, bammes_direct_2012}.

The most fragile dose sensitive samples, such as proteins, have largely been the preserve of conventional phase contrast TEM imaging as in cyro-EM. Phase contrast imaging in STEM has historically been far less dose efficient, but advanced STEM methods that also provide highly dose efficient phase contrast signals have been developed and also benefited greatly from advances in pixelated detector technology \cite{ophus_four-dimensional_2019}.  The vastly expanded information on the position dependence of the electron scattering provided in 4D STEM by capturing a 2D image of the scattering at each probe position 
provides advantages that go well beyond the much greater flexibility it provides in choosing between conventional STEM imaging modalities after taking the data \cite{pennycook_efficient_2015,hachtel_sub-angstrom_2018}. The richness of 4D STEM datasets also allow for significantly more advanced analysis. For example, center of mass (COM) and ptychographic based 4D STEM methods both provide greatly enhanced sensitivity to the local potentials of a material and greatly enhanced signal per electron compared to conventional STEM methods \cite{lazic_phase_2016,pennycook_efficient_2015, yang_efficient_2015}. Moreover ptychographic methods have been shown capable of also providing significantly enhanced dose efficiency over conventional TEM phase contrast \cite{pennycook_high_2019}, thus offering the prospect of another leap forward in low dose performance as well as a greater precision, for example for mapping local charge densities\cite{muller-caspary_atomic-scale_2018,martinez_direct_2019,madsen_ab_2021}.

4D STEM methods such as COM or ptychography do not require cameras with particularly large numbers of pixels \cite{yang_efficient_2015, muller-caspary_comparison_2019}. Given the achievable frame rate depends on the number and bit depth of the pixels due to finite readout and data transfer rates, the development of small direct electron detectors has greatly benefited 4D STEM.
Small fast hybrid pixel direct electron detectors such as the Medipix3 \cite{mir_characterisation_2017} are particularly attractive as they also offer the benefit of beam hardness, with the detection layer physically separated from the readout electronics. However, such detectors have typically been employed at frame rates of at most a few thousand frames per second \cite{ryll_pnccd-based_2016}.

By utilizing the much reduced bandwidth required for a 1-bit counting depth in each pixel of a Medipix3 detector, O'Leary et al.~demonstrated a significant speedup of 4D STEM acquisition at a 12.5 kHz frame rate, corresponding to an 80 microsecond dwell time \cite{oleary_phase_2020, nord_fast_2020}. Although this allowed them to achieve a relatively low dose of 200 e$^{-}$\AA$^{-2}$ by using an extremely low probe current in a focused probe configuration, such a dose is still much higher than required by the most beam sensitive materials~\cite{Mayoral_2015, zhu_unravelling_2017, Lozano_2018} and the speed is still far slower than the single to few microsecond dwell times used in rapid conventional ADF based STEM measurements. Detector technology continues to advance and achievable frame rates are increasing. For example 11 microsecond dwell time 4D STEM has recently been achieved with a specially designed custom extremely high frame rate camera \cite{ciston_4d_2019, ercius_4d_2020}, but this is still an order of magnitude slower than the single to very few $\mu$s typical for rapid scanning STEM, as frequently used in drift corrected time series of STEM scans..


By using a defocused probe the CBED pattern contains information from a larger area of the sample from each probe position. Although significant overlap of the illuminated sample areas from adjacent probe positions is generally required, the reduced real space sampling requirements allows defocused probe ptychography techniques \cite{song_atomic_2019,chen_mixed-state_2020}  to more easily reach low doses with relatively slow cameras compared to focused probe techniques. This advantage has recently been utilized to demonstrate defocused probe electron ptychography at extremely low cryo-EM doses with a Medipix3 \cite{zhou_low-dose_2020}, showing advantages over conventional TEM common to ptychographic techniques such as the single signed contrast transfer function (CTF) with a passband optimizable via the convergence angle and without the need for additional aberrations. However defocused probe techniques preclude the simultaneous use of the highly informative ADF based Z-contrast signal and generally involve solving for the phase using iterative convergence with the accompanying risks of incorrect convergence particularly at low doses. The focused probe single side band (SSB) \cite{rodenburg_experimental_1993, pennycook_efficient_2015} and the Wigner distribution deconvolution (WDD) \cite{noauthor_theory_1992, yang_simultaneous_2016} methods can provide simultaneous Z-contrast signals and non-iterative phase determination, and with progressing camera speedup can also now reach the extreme low dose regime of cryo-EM.  

Here we demonstrate the use of an alternative event driven detector architecture to completely remove the bottleneck in speed for 4D STEM in comparison to conventional STEM detectors based on scintillators and photomultipliers for the low probe currents desirable for low dose operation. The event driven Timepix3 detector was developed with an emphasis on time resolution for applications in a wide range of scientific disciplines with a maximum time resolution of 1.56~ns 
\cite{hatfield_lucid-timepix_2018,beacham_medipix2timepix_2011}. Event driven operation allows one to take advantage of event sparsity to achieve enhanced time resolution by avoiding the readout of pixels containing zero counts. 
In STEM one can reduce the dose imposed on a sample per unit area by reducing the probe current and dwell time, both of which increase the sparsity of the events in each probe position. 
We present results from 4D STEM performed with a Timepix3 detector at dwell times of a few microseconds down to 100~ns at both 60 and 200~kV accelerating voltages. Although the detector configuration used herein requires the use of low probe currents, we show how the speed facilitates multiply scanned 4D STEM to be used in order to increase the signal-to-noise with minimal susceptibility to drift. The results demonstrate that event driven camera technology enables the efficiency of 4D STEM and in particular electron ptychography to be exploited for both large fields of view and the extremely low doses without resorting to a defocused probe configuration.

\section{Experimental Setup}
The event based detector used in the camera setup in this work is the AdvaPIX TPX3 \cite{noauthor_advacam_nodate} which is a Timepix3 based detector where the thickness of the sensitive silicon layer is 300~$\mu$m. Timepix3 chips are 3-side tileable with a maximum event capacity of 0.43$\times$10$^6$/mm$^2$/s. A single Timepix3 chip contains $256 \times 256$ pixels where each pixel has a size of 55 $\times$ 55 $\mu$m, resulting in a theoretical capacity of 80$\times$10$^6$ counts per second for a single chip device. However due to the bandwidth of the USB 3.0 port used for the readout from our device to the computer, the maximum capacity for our single chip setup is 40$\times$10$^{6}$ counts per second when using flat field illumination. If one electron lights up one pixel, this corresponds to a current of 6.4~pA which compared to conventional STEM imaging is somewhat on the low side. In the \textit{Section \ref{sec:char}}, it is shown that this estimate has to be adjusted with more technical details but the order of magnitude is correct \cite{krause_precise_2021}. In Fig.~\ref{fig:setup} (a), a schematic view of the setup is shown where an incoming convergent electron beam interacts with the sample. 
After the interaction with the sample, the electron beam propagates into the far field regime where the detector is placed. In Fig.~\ref{fig:setup} (b) an example of the data is shown where for every incoming event its point of impact and time of arrival (TOA) are saved. Knowing the probe position at each time, the conventional 4D STEM dataset can be obtained. In Fig.~\ref{fig:setup}(c) some measured diffraction patterns are shown in which the sparsity is seen due to the low dose per scanned point. The resulting position averaged convergent beam electron diffraction (PACBED), calculated from the sum of all diffraction patterns, is also shown.

The Timepix3 detector was mounted on a probe-corrected FEI Themis Z using a custom designed retractable mount interface shown in Fig.~\ref{fig:setup} (d). In order to synchronize the scan coils with the detector, a custom scan engine is used \cite{zobelli_spatial_2019,noauthor_attolight_nodate}. The detector and scan engine are synchronized by using a 10~MHz reference clock output which is then multiplied with a custom designed phase locked loop circuit to act as a 40 MHz master clock for the Advapix in order to keep both scan engine and camera synchronised in time. At the start of the acquisition, a synchronisation signal is generated by the scan engine which triggers a recording sequence on the detector side.
In this work, no flyback time is applied since we aim to reduce the dose as much as possible and we do not have access to a fast enough beam blanker to shut the beam down. Both the detector and scan engine have an application programming interface (API) in Python3 giving us the ability to automate the acquisition sequence. The incoming raw data from the detector is processed using self written software and the ptychographic reconstruction is performed using the single side band ptychography reconstruction algorithm. The full recorded dataset and data processing software is made available in Zenodo allowing others to duplicate our efforts or to experiment with new algorithms that are optimised for event based data streams \cite{jannis_event_2021}.

\section{Timepix3 Characterization}\label{sec:char}
\begin{figure*}[!htb]
\includegraphics[width=\textwidth]{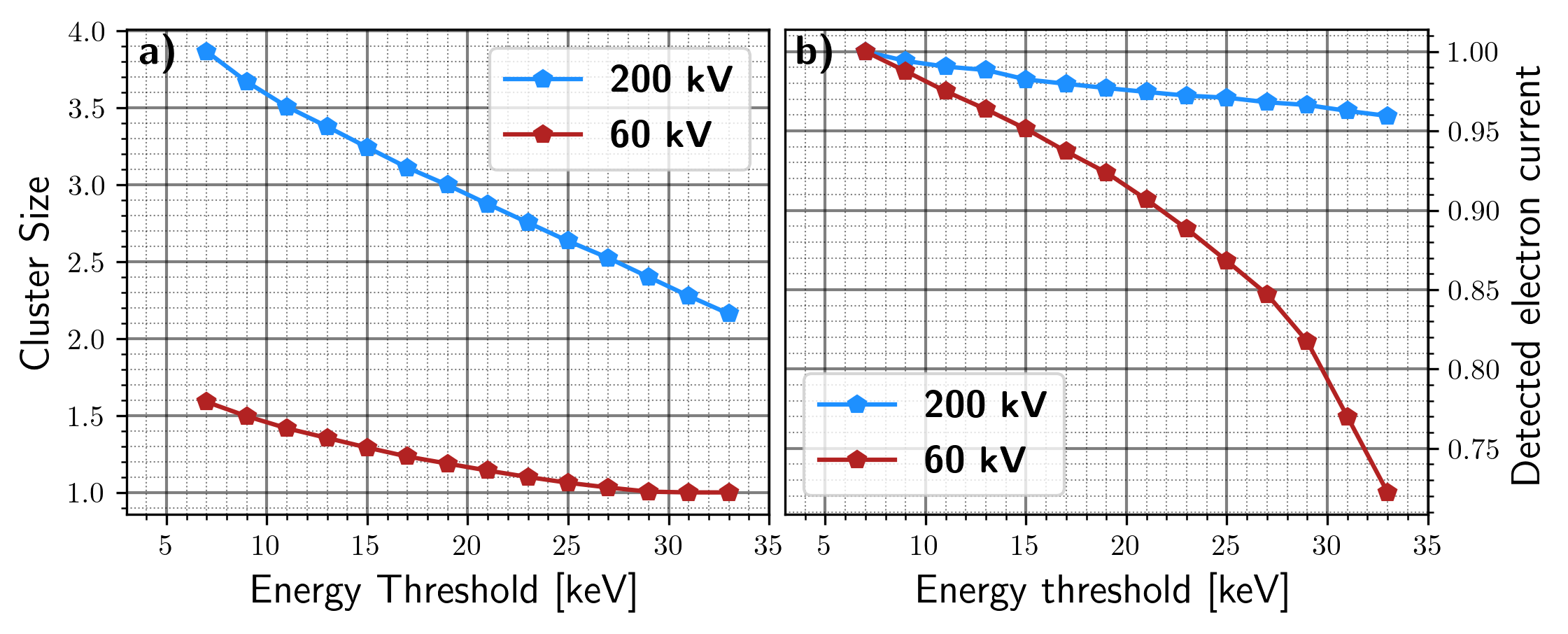}
\caption{\label{fig:char_tpx3} \textbf{(a)} The cluster size as a function of energy threshold for the two acceleration voltages 60 and 200~kV. \textbf{(b)} The detected electron current which is calculated by dividing the number of events per second on the detector by the average cluster size. The detected current is relative compared to the detected current at a energy threshold of 7~keV.
}	
\end{figure*}

The Timepix3 detector is a hybrid direct electron detector where the silicon sensor layer is bonded onto the underlying processing ASIC electronics similar to the Medipix chips \cite{rosenfeld_editorial_2020, ballabriga_asic_2018}. As for the Medipix detectors, the main parameter to be changed is the energy threshold which is a pre-set discrimination level used to discriminate an electron event above the noise floor. Hence, by varying this parameter we can select the amount of signal detected for a particular beam current. The energy threshold influences mainly two things, firstly the amount of detected events when one electron hits the detector and secondly the number of detected electrons. Throughout the rest of the paper the amount of pixels which are excited by a single electron will be referred to as a cluster. The identification of the clusters is derived from the work by van Schayck \textit{et al.} \cite{van_schayck_sub-pixel_2020} where clusters are identified by checking if neighbouring pixels are excited within a small time interval ($\sim$100~ns). The energy threshold also influences the number of detected electrons because increasing the threshold will results in some events remaining undetected since they do not exceed the threshold value when scattering inside the sensitive silicon layer.

The camera used in this work shows a workable calibrated threshold range between 7 and 33~keV, in \textit{Supplementary Material S1} flat field images of different threshold values are shown where the artefacts for lower and higher threshold values are indicated. Note that these settings depend on an internal calibration routine that is performed when producing the detector \cite{jakubek_2011, kroupa_2017, urban_2021}. During the experiments it was observed that the temperature of the detector chip remained constant around 30$^{\circ}$.

In Fig.~\ref{fig:char_tpx3} (a), the cluster size as a function of energy threshold is plotted for 60 and 200~kV acceleration voltages. As expected, the cluster size decreases as a function of increasing energy threshold where for 200~kV the lowest cluster size is approximately 2.2 pixels and for 60~kV the size drops to 1 pixel at a threshold of 33~keV. In Fig.~\ref{fig:char_tpx3} (b), the detected electron current is shown as a function of energy threshold for the two acceleration voltages. The detected electron current is calculated by dividing the number of events on the detector by the cluster size within a fixed amount of time (two seconds in this experiment). From Fig.~\ref{fig:char_tpx3} (b), it is seen that at higher energy thresholds (33~keV) for 60~kV, a significant amount of electrons (25 \% at 33~keV) remain undetected. 

Since the incoming electron current is lower than the detection threshold for measuring the current via the fluorescent screen or spectrum drift tube method, it was not possible to measure the current during the characterization experiments. Therefore, an additional measurement was performed to verify that the detector is not operated in an undercounting regime. This was done by using a parallel beam with sufficient current (30~pA) such that the fluorescent screen is able to measure it. Next a small aperture was inserted from which the remaining current can be estimated via the radii of both the original beam and the apertured beam assuming a homogeneous current density in the original beam. The apertured beam is then placed on the detector from which the number of electrons detected can be measured. A threshold of 7~keV is used to be sure that the least amount of electrons are lost. The result is shown in Table~\ref{tab:current} where it is observed that the measured extrapolated current from the fluorescent screen and the Timepix3 are very similar indicating that the Timepix3 was not operated in undercounting mode. The error on the current measurement from the fluorescent screen is approximated to be around 10\%~\cite{krause_precise_2021}. A faraday cup could be used to get a more precise measurement on the actual current but was unavailable for our experiments. Moreover in the work of Krause~\textit{et al.}  \cite{krause_precise_2021} it is also shown that direct electron detectors provide an accurate estimate for low currents in range of acceleration voltages used during the experiments. 

\begin{table}
\centering
\begin{tabular}{|c|c|c|}
	\hline
	&Extrapolated fluorescent & Timepix3   \\ 
	&screen current &  current (7~keV)  \\ \hline
	
	60~kV& 0.1$\pm$0.01~pA  & 0.1~pA   \\ \hline
	200~kV& 0.1$\pm$0.01~pA &  0.09~pA \\ 
	\hline
\end{tabular}
\caption{The measured extrapolated current on the fluorescent screen and on the Timepix3 operated at a threshold of 7~keV. This small current was measured on the fluorescent screen by first measuring the current of a large parallel beam. Afterwards, a small aperture is inserted to block a large part of the beam. The current through the aperture can be calculated from knowing the radii of both the parallel beam and aperture. The error on the extrapolated fluorescent screen is approximated to be 10\%.}\label{tab:current}
\end{table}

Since the 7~keV threshold loses the least amount of electrons (see Fig.~\ref{fig:char_tpx3}), this threshold is the most accurate measurement of the incoming electron count rate. Note that this is a lower boundary on the actual current since some electrons will backscatter and other electrons will scatter inside the silicon layer without exceeding the threshold energy.
Next to the backscattering of the incoming electrons, electron counts can be lost due to the dead time of the individual pixels which is $\sim$500~ns. For the flat field illumination acquisition, the detected incoming electron count rate is 14$\times$10$^6$ counts/s (2.4~pA) and 6$\times$10$^6$ counts/s (1~pA) for respectively 60 and 200~kV at a threshold of 7~keV. Therefore the probability for an electron to arrive in the same pixel area within 500~ns is in the order of 0.01\% which is a very small fraction. When the illumination area decreases or the incoming current increases, the electrons missed due to the finite dead time of the individual pixels will increase and can become measurable~\cite{jannis_2021}. In the rest of this work, the total incoming electrons are calculated by dividing the number of detected events by the cluster size which is then multiplied by the ratio between the  electron current at 7~keV and the used threshold value (see Fig.~\ref{fig:char_tpx3}). \\
For example if 1$\times$10$^6$ events per second are detected with a threshold of 30~keV at a 60~keV accelerating voltage, the incoming electron current is calculated by multiplying this value by 1/1.02, the inverse of the cluster size, and 0.79, the ratio between the detector electron current at 7 and 30~keV. \\
From Fig.~\ref{fig:char_tpx3} we can conclude that at this lowest threshold, more electrons are detected but they make on average a larger cluster size reducing the maximum current for which reliable detection and readout can be achieved. At 200~kV, the amount of events at the highest threshold (33~keV) is only 4\% less compared than when using a threshold of 7~keV while it significantly reduces the cluster size. This results in a higher allowable beam current as compared to lower threshold levels since there is an upper limit on the maximum count rate of the detector. \\
The detector was also briefly tested with an acceleration voltage of 300~kV. The cluster size increased quite considerably to an average of six pixels at a threshold of 33~keV. This meant that the incoming current should be dropped by a factor of three compared to 200~kV which makes the use of 300~keV electrons much less attractive and is therefore not considered further. \\
When increasing the threshold it is expected that the detected counts do not arise exactly at the position where the electron enters since the electron loses most of its energy towards the end of its track. Hence even when the average cluster size is one, it is expected that the modulation transfer function (MTF) will not approximate the theoretical limit of a square pixel ~\cite{paton_2021}. The identification of clusters can be used to increase the MTF of the detector as shown in the work of van Schayck \textit{et al.} \cite{van_schayck_sub-pixel_2020} where they trained a neural network to better estimate the point of initial impact. However for our work such increasing of the MTF will not improve the results from our 4D STEM measurements since the algorithms used to reconstruct the signal are not sensitive to this \cite{yang_efficient_2015, muller-caspary_comparison_2019} .
In \textit{Supplementary Material S2}, the corrected electron signal, where the point of impact is calculated using mean x and y coordinate of the cluster, is used to investigate if the integrated center of Mass (iCOM) result changes due to this improvement of MTF. In \textit{Supplementary Material S2} it is shown that a small discrepancy of approximately 1\% between both reconstructed signals is observed. Therefore, when determining the positions of atoms and defects is the goal, declustering is not necessary, but when the aim is to obtain a quantitative iCOM signal then declustering is likely desirable.
During the experiments, it was noted that the detector does not record the incoming signal for short periods of time ranging from 50~$\mu$s to 1.75~ms. This happens for $\sim$0.12\% of the time that the detector is recording. This artefact is shown in \textit{Supplementary Materials S3}.  Up until now the origin of this artefact is not known but it is not problematic since it only occurs for such a small amount of time. In this work we took the simple approach of replacing the integrated signal of a particular probe position, where the detector is not recording, by the average signal. In the future better ways of reconstructing these signals can be performed, and hopefully the losses fully avoided in the first place, but this is outside the scope of this work. Since the amount of lost electrons is relatively small, this artifact does not significantly hinder the performance of the camera.

\section{Results \& Discussion}
\begin{figure*}[!htb]
\includegraphics[width=\textwidth]{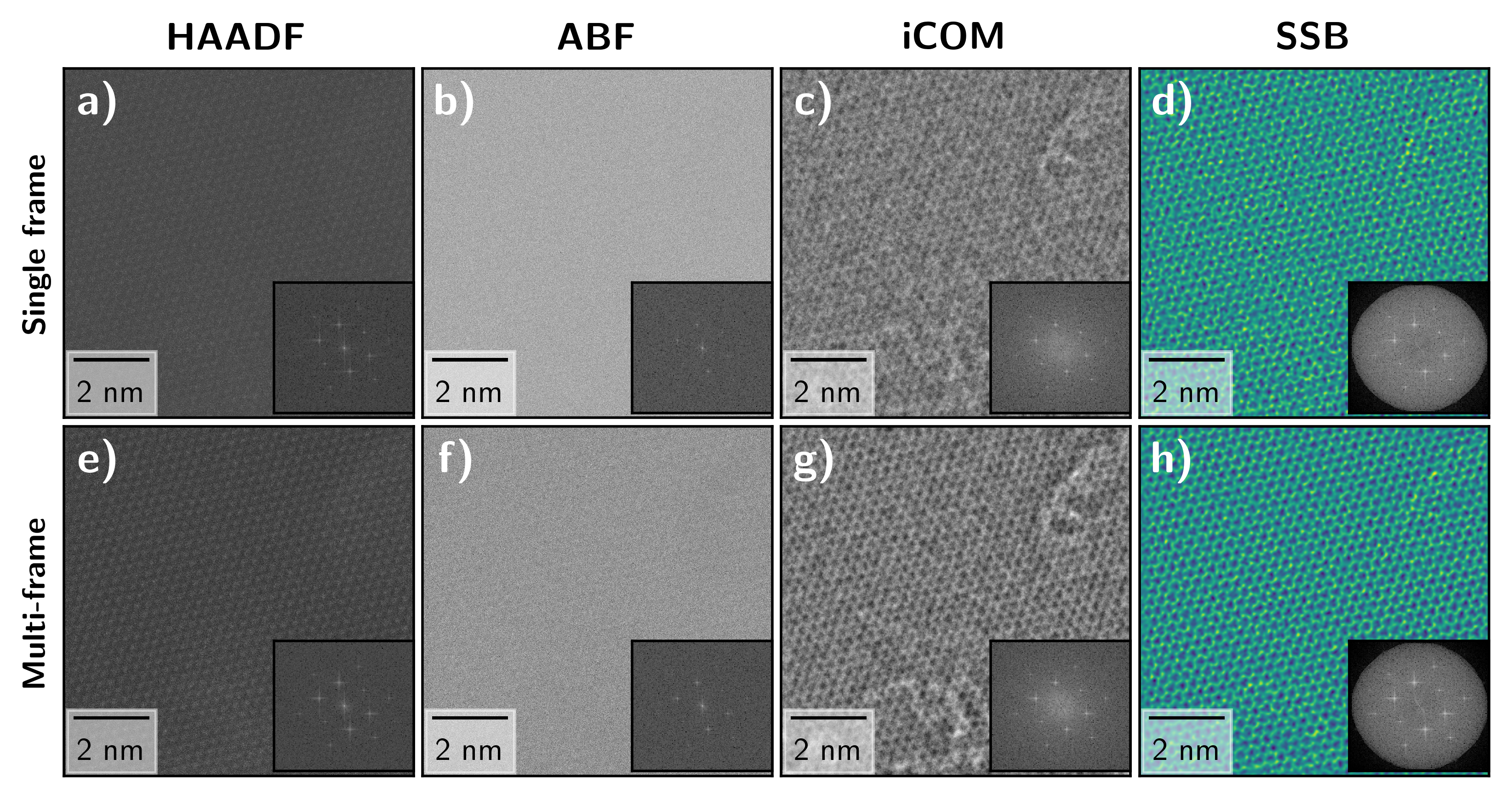}
\caption{\label{fig:ws2_overview} \textbf{(a,e)} Single and summed multi-frame HAADF images from a $3 \times 1024 \times 1024$ scan at 6~$\mu$s dwell time. The acceleration voltage used during the acquisition is 200~kV. The sample is a monolayer of WS$_2$. The HAADF signal is collected with the conventional HAADF detector. \textbf{(b,f)} Single and multi-frame ABF images reconstructed from the simultaneously acquired 4D STEM dataset. \textbf{(c,g)} The reconstructed iCOM images from the single and multi-frame scans. \textbf{(d,h)} The reconstructed SSB images from the single and multi-frame scans.    
}	
\end{figure*}

\begin{figure*}[!htb]
\includegraphics[width=\textwidth]{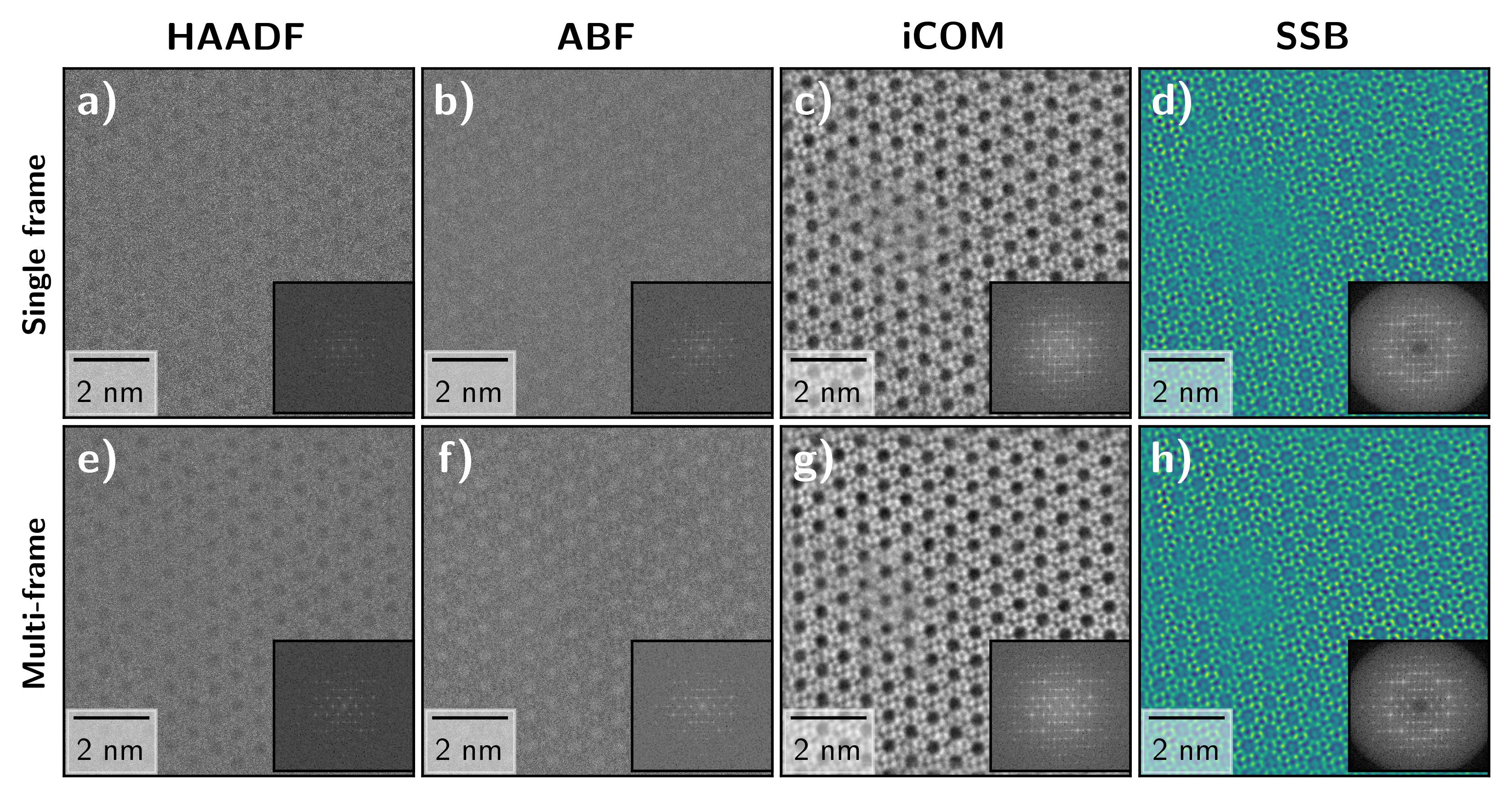}
\caption{\label{fig:si1_overview} \textbf{(a,e)} Single and summed multi-frame HAADF images from a 1~$\mu$s dwell time $10 \times 1024 \times 1024$ scan of a silicalite-1 zeolite sample. The acceleration voltage used during the acquisition is 60~kV.\textbf{(b,f)} Single and summed multi-frame ABF images reconstructed from the use of virtual detector. \textbf{(c,g)} The reconstructed iCOM images from the single and multi-frame scans. On the left of the image the distortions due to finite response time of the scan coils are visible. \textbf{(d,h)} The reconstructed SSB images from the single and multi-frame scans.    
}	
\end{figure*}

\begin{figure*}
\includegraphics[width=\textwidth]{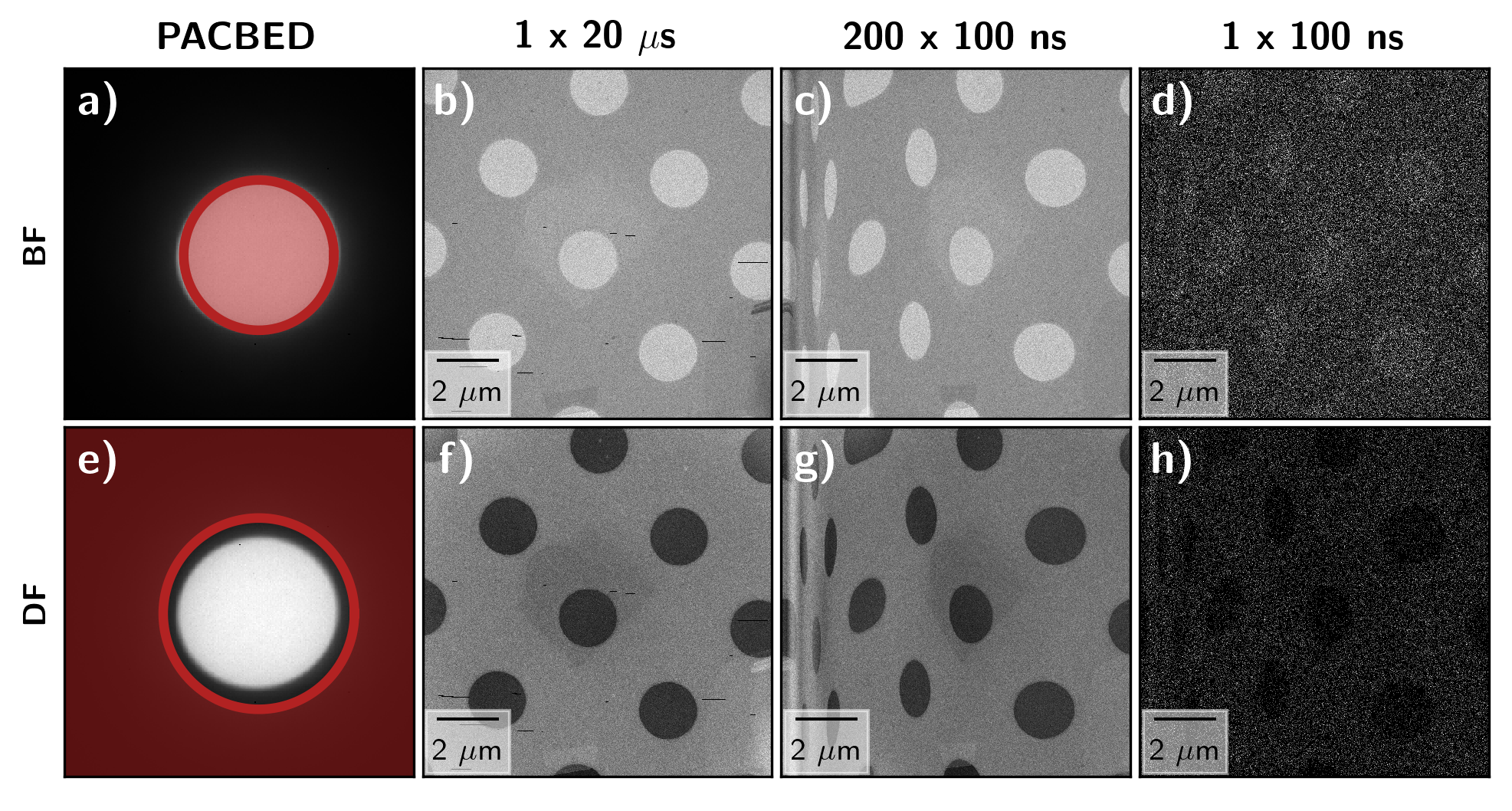}
\caption{\label{fig:fast_scan} \textbf{(a,e)} The PACBED of the 4D STEM scan where, for both signals bright field (BF) and dark field (DF), the virtual detectors are shown. \textbf{(b,f)} The BF and DF signal of a $1024 \times 1024$ scan at 20~$\mu$s dwell time where the sample is a low magnification image a holey carbon film. The acceleration voltage used during the acquisition is 60~kV. \textbf{(c,g)} The summed BF and DF image of the same scan size with a dwell time of 100~ns and 200 frames which are scanned. Clear distortions arising from the response time of the coils are visible but the detector has no significant issues with this dwell time. \textbf{(d,h)} A single BF and DF frame of the fast scan using a dwell time of 100~ns. 
}	
\end{figure*}

A 2D WS$_2$ sample was used to showcase the performance of the detector at 60~kV. A multi-frame acquisition is performed where the probe, with a convergence angle of 25~mrad, is scanned over $1024 \times 1024$ probe positions three times at a dwell time of 6~$\mu$s. The dose per frame is estimated to be 6000 e$^{-}$\AA$^{-2}$ . The size of the dataset is 6.6~GB which is a significantly lower data storage requirement than using a 1-bit mode of an equivalently sized frame-based detector where the data size would be 33~GB. \\
In recent years, another type of data storage for frame based detector has been developed. The compression method is named electron-event representation (EER) which stores each electron-detection event as a tuple of position and time\cite{guo_electron-event_2020}. This is similar to the output of event based detectors. However the speed of the frame-based camera is still determined by the fps which is not the limit for the event based detectors \cite{datta_data_2021, pelz_real-time_2021}.\\
During the acquisition, the signal from the HAADF detector was read out simultaneously to collect the electrons which scatter to higher angles than the pixelated detector and provide a simultaneous Z-contrast image. In this work we provide three different signal reconstruction methods which are annular bright field (ABF), where only events which land on a annular ring on the detector is integrated from 33 to 100\% relative to the convergence angle in order to obtain the image. Secondly, the iCOM signal is reconstructed from calculating the centre-of-mass of each probe position on which a two dimensional integration is performed which yields a scalar image. Finally, the single side-band (SSB) method is used to retrieve the phase of the object. The ABF and iCOM signals are calculated directly from the event list (see Fig.~\ref{fig:setup}(b)) whereas for the SSB reconstruction the data was converted to standard non-sparse four-dimensional data for the most facile input to our existing ptychographic processing software, which was written with framing cameras in mind. 
In the future the algorithms can be modified to accept the sparse event data directly or even perform `live' imaging \cite{pelz_real-time_2021}. 
The resulting signals (ABF, iCOM and SSB) created from the 4D STEM images are shown in Fig.~\ref{fig:ws2_overview}. The HAADF signal in Fig.~\ref{fig:ws2_overview}(a,e) arises from the conventional HAADF detector (47-210 mrad) which we recorded in parallel with the event based detector data. The reconstructed multi-frame images are aligned using an open-source software which was developed for low signal-to-noise cryo-STEM data \cite{savitzky_image_2018}. The software uses all possible combinations of image correlations, instead of using a single reference image, to determine the optimal shift. The relative image distortions between the scans are calculated using the SSB images due to its good contrast. The resulting images shifts were then also applied to the HAADF, ABF and iCOM images to align them. In Fig.~\ref{fig:ws2_overview}(b,f) the ABF signal is shown where the low contrast arises as a result from the low dose conditions. This is not particularly surprising as in the weak phase object approximation (WPOA) zero contrast is expected when using a centrosymmetric detector configuration, and the 2D WS$_2$ is of course thin, relatively weak and lacking in channeling contrast in comparison to typical 3D materials \cite{pennycook_efficient_2015}.

The iCOM signal shown in Fig.~\ref{fig:ws2_overview}(c,g) and the SSB ptychographic reconstruction shown in Fig.~\ref{fig:ws2_overview}(d,g) both take greater advantage of the 4D data and show much stronger signals.
The difference in contrast of the SSB with respect to the iCOM signal arises from the different contrast transfer functions (CTF) of both signals \cite{lazic_phase_2016, oleary_contrast_2021} and the way in which the ptychographic method keeps track explicitly of where in probe reciprocal space each frequency is transferred \cite{pennycook_efficient_2015, yang_efficient_2015}. The lower limit of the electron dose is calculated using the information from Fig.~\ref{fig:char_tpx3} (b) where the experiment was performed at a energy threshold of 30~keV indicating that minimally $\sim$23\% of the incoming electrons are not detected. 
This gives a minimum dose of $\sim$6000 e$^{-}$\AA$^{-2}$  per frame where the incoming electron current used was 4.4~pA.  The real dose used during the acquisition would be slightly higher since there is a small fraction of the electrons which is not detected due to the backscattering and pixel dead time. However as derived from Table~\ref{tab:current}, it is expected that the deviation between detected events and actual beam current is $\pm$~10\%.

The same type of data acquisition at 60~kV can be performed at 200~kV where the main difference is the larger cluster size at 200~kV. However in terms of lost electrons, the 200~kV is better since only a small fraction (3\%) seems to be lost by increasing the energy threshold to the 31~keV used for the 200~kV data presented here taken of a silicalite-1 zeolite \cite{fujiyama_entrance_2015}. A $10 \times 1024 \times 1024$ probe position scan is performed at a dwell time of 1~$\mu$s where the dose per frame was 300 e$^{-}$\AA$^{-2}$  and the total dose over the ten scans was 3000 e$^{-}$\AA$^{-2}$. The current used during the acquisition was 1.2~pA. The convergence angle used during the experiment was 12~mrad. In Fig.~\ref{fig:si1_overview}, the reconstructed signals are shown where the HAADF and ABF shows, as expected, very low contrast. As for the  WS$_2$, the iCOM and SSB signals both show good contrast.
The distortions visible on the left of the images are due to the slow response of the scan coils at this dwell time as we have no flyback time.\\
To check if we were able to record 4D STEM scans at dwell times of 100~ns and that the detector is still able to record a proper 4D STEM dataset, a low magnification scan on a holey carbon film was performed. This sample and magnification was selected to have a high contrast and to minimize the effect of drift since the individual 100~ns scans have such a low signal-to-noise that aligning them would be very challenging and beyond the scope of the present study. First a $1024 \times 1024$ scan at a 20~$\mu$s dwell time was performed, the bright field (BF) and dark field (DF) images are shown in Fig.~\ref{fig:fast_scan}(b,f). The virtual detectors used to reconstruct the signal are shown in Fig.~\ref{fig:fast_scan}(a,e) where the BF signal arises from the integration over the entire central beam and the DF collects all the signal which lands outside the central beam. Further the same field of view is scanned with the same number of probe positions except the dwell time is decreased to 100~ns where a total of 200 frames were scanned to have an equal dose as the slower 20~$\mu$s scan. The resulting summed BF and DF are shown in Fig.~\ref{fig:fast_scan}(c,g). Large distortions due to the finite response of the  scan coils are visible but in essence, the 4D STEM signal is recorded properly showing that the detector can handle such short dwell times with ease. In Fig.~\ref{fig:fast_scan}(d,h) the BF and DF of a single scan are shown where the low signal-to-noise results from the very small number of counts per probe position.
These results show that dwell times in the order of 100~ns would be possible if the scanning system bandwidth could be improved, possibly at the expense of maximum field of view capabilities. Efforts in this direction are reported for magnetic deflection \cite{ishikawa_high_2020} but also electrostatic deflection coupled to high bandwidth amplifiers are a possibility.
Beside the practical limitations of the scan system, it is unlikely that we can reduce the dwell time towards the ultimate timing resolution from the Timepix3 being 1.56~ns as synchrotron experiments have shown that significant timing inaccuracies can arise increasing this up to $\sim$20ns~\cite{Crevatin_2016}. Nevertheless for any practical TEM setup, the scan engine will currently be the limiting factor by far.\\
Due to the speed at which full 4D STEM data sets can now be recorded with event based detectors, this technique has the capability to be combined with other conventional high resolution STEM (HR-STEM) methods. For instance, tomographic series can be performed where instead of just the HAADF signal, the full diffraction pattern can be recorded where a large range of scattering angles is available. 
In the post-processing steps, different signals can be reconstructed to increase the information gathered from the experiment \cite{chang_ptychographic_2020} with the potential for even online processing of the datastream \cite{clausen_libertemlibertem_2021}.\\
Another method that becomes more attractive for 4D STEM is depth sectioning where instead of the simple fixed focus multi-frame acquisition used in this work, between each frame the focus is changed in order to get three dimensional information about sample \cite{van_benthem_three-dimensional_2005, ishikawa_single_2016,xin_aberration-corrected_2009, pennycook_3d_2017,yang_imaging_2015}. Adding the 4D STEM dataset to such optical sectioning could potentially improve its performance where for instance for S-matrix reconstruction this depth sectioning is necessary \cite{pelz_reconstructing_2020}. Finally, acquiring 4D STEM data when the scan sequence is changed is easily accessible if one knows at each time where the probe is positioned. These different scanning strategies are used e.g. to decrease damage \cite{velazco_reducing_2021, nicholls_minimising_2020}, or distortions \cite{sang_dynamic_2016, velazco_evaluation_2020}.\\
Although the main limit of the current Timepix3 camera is its limited detectable count rate, new plans for a Timepix4 chip have been revealed where the number of pixels almost quadruples to $512 \times 448$ and the maximum detectable count rate increases by a factor of six. This allows for a current increase by approximately a factor of 24, bringing it,  in the coming years, much more in line with conventional beam currents used in STEM imaging \cite{xavier_llopart_timepix4_2020}.

\section{Conclusion}
In this work, it is shown that using a hybrid pixel direct electron event based detector rather than conventional frame-based cameras enables the recording of 4D STEM datasets with dwell times as low as 100~ns. The detector was characterised at two different acceleration voltages, 60 and 200~kV. For 200~kV, the maximum electron count rate is half that of 60~kV. However when increasing the threshold for 60~kV, a significant decrease in collection efficiency is observed. Hence when performing experiments, a compromise on the threshold is required where the higher the threshold, the higher the electron dose rate that can be detected. However this decreases the collection efficiency which is detrimental for beam sensitive materials. Furthermore by synchronizing the detector with a versatile scan engine, multi-frame 4D STEM acquisitions could be performed with scan sizes of $1024 \times 1024$ or larger using dwell times in the order of $\mu$s. This opens up the possibility to always perform 4D STEM acquisition instead of conventional STEM and profit from the significantly higher information content about the sample for the same incoming beam current. In the future, we anticipate improvements in both the data processing times and the maximum count rates detectable by event based detectors to make this type of setup more straightforward. 
\section*{Acknowledgements}

This project has received funding from the European Union's Horizon 2020 Research Infrastructure - Integrating Activities for Advanced Communities under grant agreement No 823717 – ESTEEM3. J.V. and A.B. acknowledge funding from FWO project G093417N (`Compressed sensing enabling low dose imaging in transmission electron microscopy'). J.V. and D.J. acknowledge funding from FWO project G042920N `Coincident event detection for advanced spectroscopy in transmission electron microscopy'. We acknowledge funding under the European Union’s Horizon 2020 research and innovation programme (J.V. and D.J under grant agreement No 101017720, FET-Proactive EBEAM, and C.H., C.G., X.X. and T.J.P. from the European Research Council (ERC) Grant agreement No. 802123-HDEM).

All data discussed in this manuscript is openly available through Zenodo to stimulate further research on this topic and to improve reproducibility.

\bibliography{TPX3_PAPER_DJ}

\pagebreak
\onecolumn

\section{Supplementary Material}
\section*{S1. Energy threshold variation for flat field illumination} \label{sec:etvfffi}
\begin{figure}[h]
	\centering
	\includegraphics[width=0.65\textwidth]{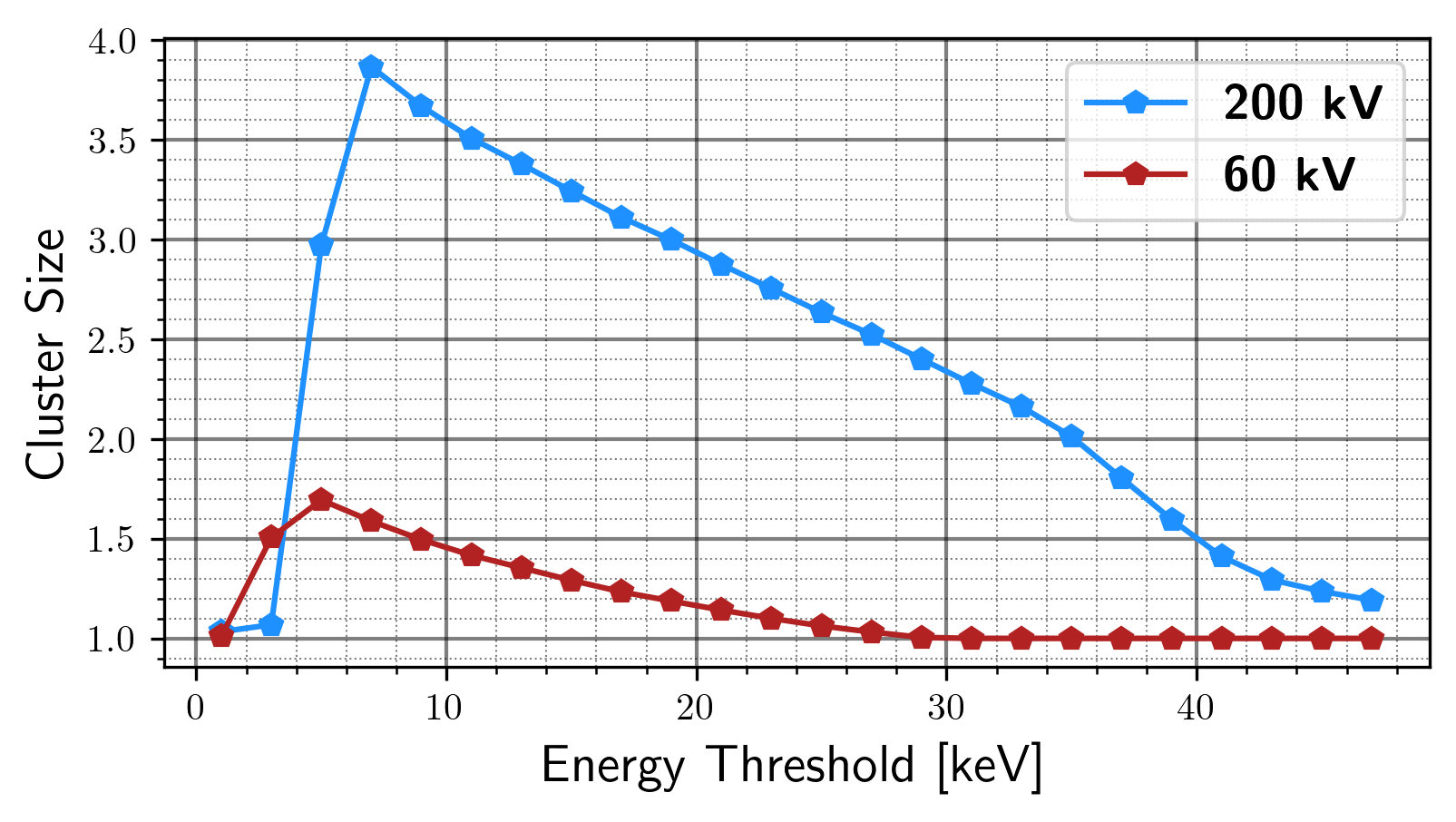}
	\caption{\label{fig:clust} The average cluster size calculated as a function of threshold for 60 and 200~kV accelerating voltages.
	}	
\end{figure}

\begin{figure}[h]
	\includegraphics[width=1.0\textwidth]{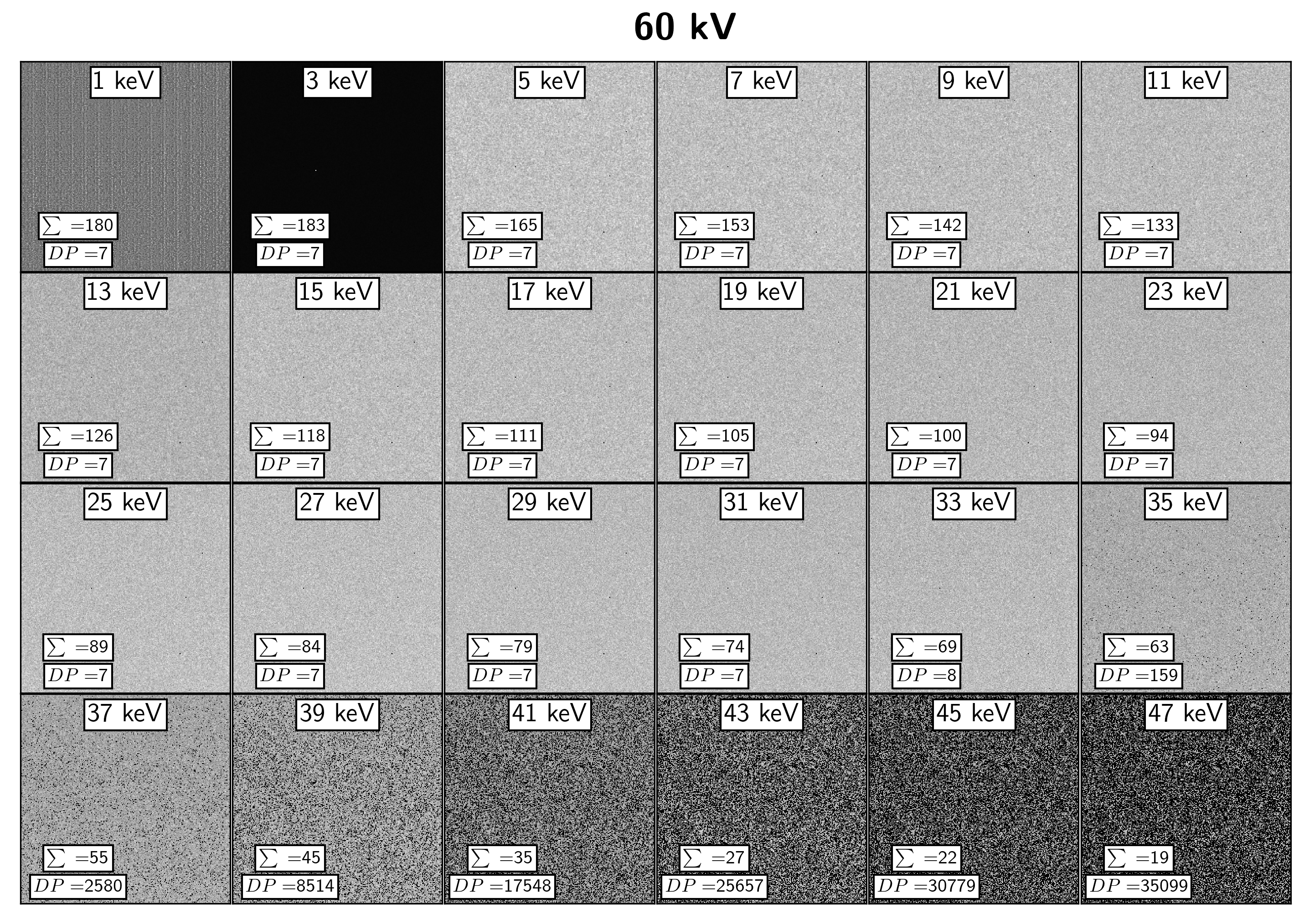}
	\caption{\label{fig:char_60} The flat field illumination of 60~kV electrons where the threshold is varied. The relative total number of detected electrons compared to 21~keV in percentage is shown and also the number of dead pixels is indicated in the plots. 
	}	
\end{figure}

\begin{figure}[h]
	\includegraphics[width=1.0\textwidth]{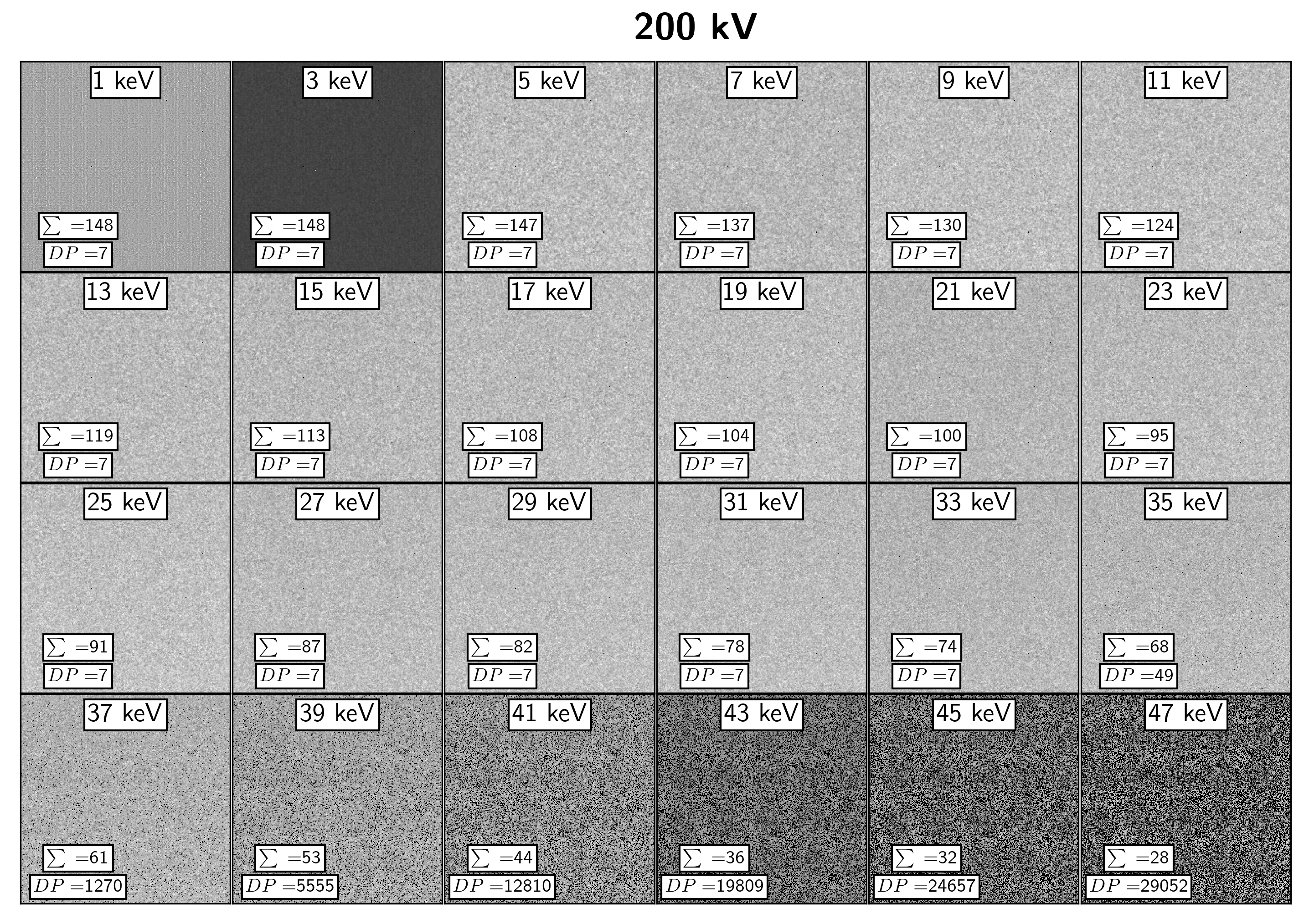}
	\caption{\label{fig:char_200} The flat field illumination of 200~kV electrons where the threshold is varied. The relative total number of detected electrons compared to 21~keV in percentage is shown and also the number of dead pixels is indicated in the plots. 
	}	
\end{figure}

In Fig~\ref{fig:clust}, the average cluster size as a function of threshold is shown. At low thresholds from 1-5~keV the cluster size decreases with decreasing threshold. This might seem nonphysical since the lower the threshold the larger the cluster size induced by an electron hit should be. However, one possible explanation for this effect would be that at low thresholds $\leq$5~keV thermal noise or other types of random noise start to be detected resulting in a smaller average cluster size since the average cluster size of random noise is one, although we note that at 1~kV the grid pattern artefact that appears in the flat field images is likely due to the underlying detector geometry. The results of flat field illumination of the incoming electron beam at 60 and 200~kV accelerating voltages are shown in Fig. \ref{fig:char_60} and \ref{fig:char_200} as a function of the threshold voltage. In order to remove the influence of such noise, the minimum usable threshold was determined to be at 7~keV. From Fig.~\ref{fig:char_60} and \ref{fig:char_200}, a maximum value of the threshold is determined by investigating the number of dead pixels as a function of threshold. A dead pixel is defined as a pixel which is not giving any counts during the acquisition. At 35~keV a sharp increase in the number of dead pixels is seen making higher threshold values undesirable for the experiments. Therefore the valid range of thresholds is determine to be between 7 and 33~keV. 


\section*{S2. Raw vs declustered 4D STEM dataset} \label{sec:rvc}
\begin{figure}[!htb]
	\centering
	\includegraphics[width=1.\textwidth]{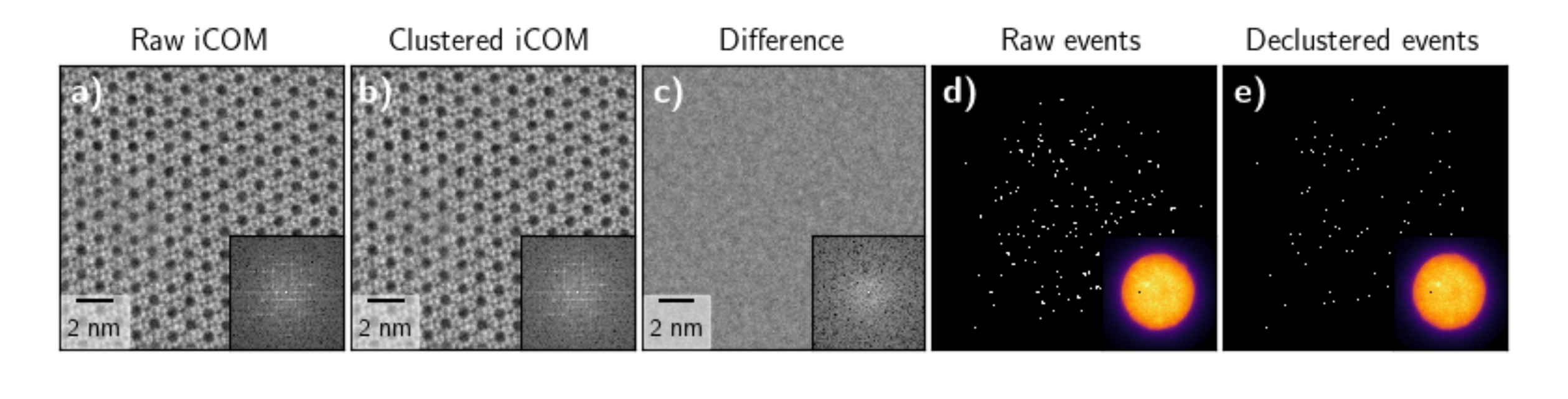}
	\caption{\label{fig:raw_vs_cl} \textbf{(a)} The reconstructed iCOM image of the raw dataset from a silicalite-1 zeolite. The scan size is $1024 \times 1024$, the dwell time is 6~$\mu$s, and the acceleration voltage is 200~kV. In the insets the Fourier transform of the iCOM signal is shown. \textbf{(b)} The same as (a) except declustering is applied.  \textbf{(c)} The difference between (a) and (b) showing the similarity between the two signals. \textbf{(d)} a set of 500 sequential events from the same dataset detected on the Timepix3 camera, in which the declustered event is clearly seen. In the inset the PACBED pattern is shown. \textbf{(e)} The same as (d) but now with declustering applied. 
	}	
\end{figure}

In this section, we investigate if the clustering of the electron events has a significant influence on the reconstructed integrated centre-of-mass (iCOM) signal. An electron probe was scanned with a 6~$\mu$s dwell time over $1024 \times 1024$ points of a zeolite silicalite-1 sample at a 200~kV accelerating voltage. The declustering algorithm searches for adjacent pixels which are excited within a time interval of 100~ns \cite{van_schayck_sub-pixel_2020}. From this cluster the new corrected time-of-arrival (TOA) is when the first pixel of the cluster is excited. The point of impact is calculated using the centre-of-mass of the cluster. In Fig. \ref{fig:raw_vs_cl} (a,b), respectively the raw and declustered reconstructed iCOM images are shown and seen to be very similar. The difference between the two signals is shown in Fig.~\ref{fig:raw_vs_cl}~(c),  showing that for iCOM the reduction of modulation transfer function (MTF) has no visual influence on the result. The average value of the difference is 1\%. Hence when the iCOM signal is used to extract atomic positions, declustering will not greatly improve the data quality. However, when quantitative information is desired from the iCOM signal, declustering is expected to improve the results. In Fig.~\ref{fig:raw_vs_cl}(d), 500 events collected sequentially from the dataset are shown from which the clusters are clearly visible. In (e), the corrected events are shown with the estimated points of impact of the electrons. In the inset figures the position averaged convergent beam electron diffraction (PACBED) of both raw and declustered datasets are shown in which also no clear difference is observed. 

\section*{S3. Artefacts in Timepix3 data}\label{sec:artifact}
\begin{figure}[!htb]
	\includegraphics[width=\textwidth]{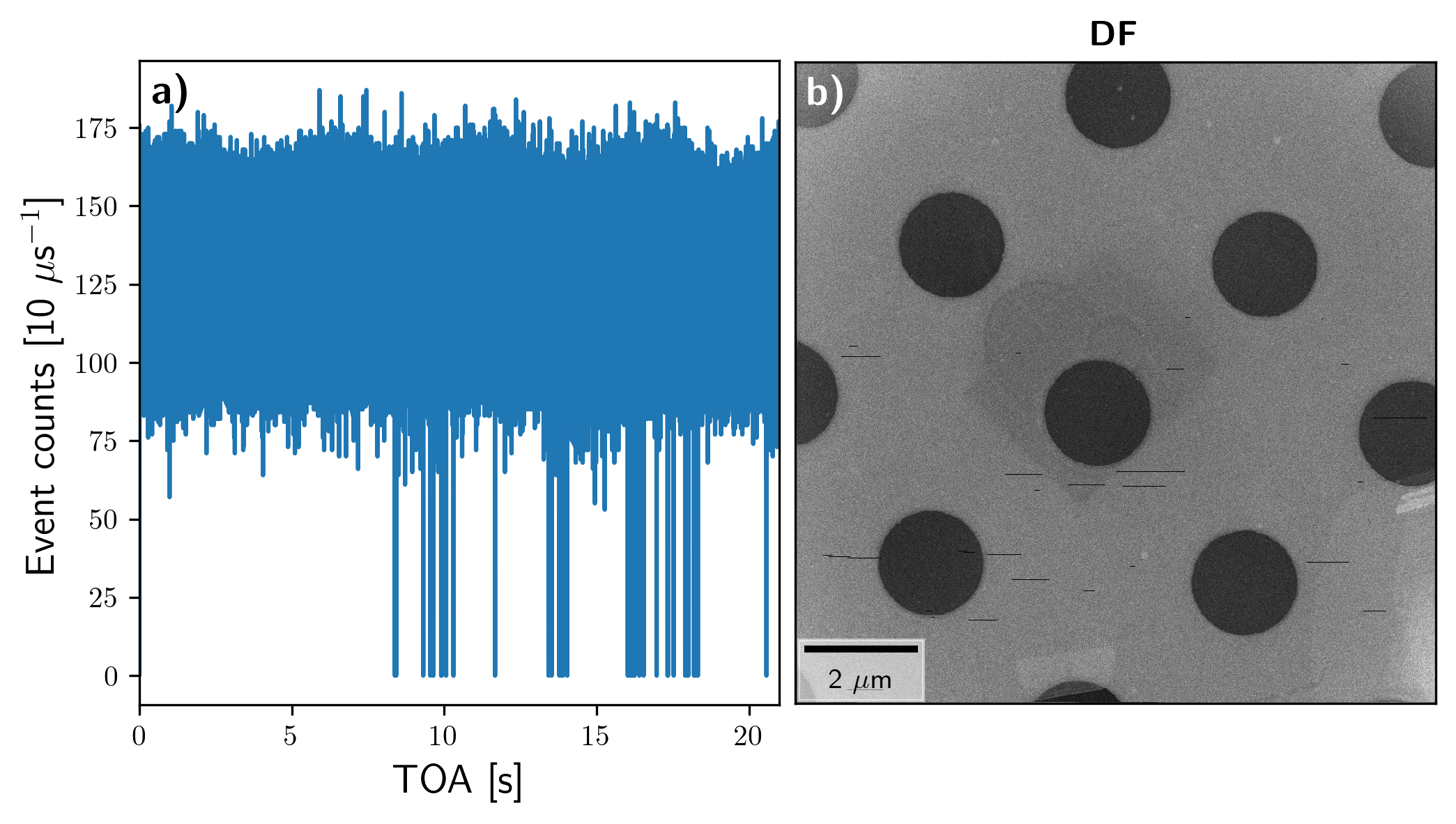}
	\caption{\label{fig:artifact} \textbf{(a)} A histogram of the TOA of the incoming events with a bin size of 10~$\mu$s. \textbf{(b)} The reconstructed DF signal where the dark lines correspond to the artefacts which arise when the detector does not detect signal in some time windows. 
	}	
\end{figure}

When recording a data stream we find that sometimes the camera does not record the incoming electrons for certain time periods ranging from 50~$\mu$s to 1.75~ms. To investigate this, the dataset of the low magnification scan used for Fig.~5 of the main text is used. In Fig.~\ref{fig:artifact} (a), a histogram calculated from the TOA of the incoming events with a bin size of 10~$\mu$s is shown. This shows the number of events where at some particular times, the counts drop unphysicaly to zero. The time that the camera is down for this acquisition is 0.12\%, which is a relatively small proportion of time compared to the entire acquisition. In Fig.~\ref{fig:artifact} (b), the corresponding dark field (DF) image is shown where the virtual detector region is shown in Fig.~5. The black lines on the image are due to the artefacts. In this work a very simple method is used to mitigate the artefacts by filling these pixels with the average value of the image. While this is likely not the best way to reconstruct the images, this is beyond the scope of our present work. In the future we expect other improved methods for image restoration or ideally ways to avoid the artefacts in the first place can be developed.  

\end{document}